\documentclass[conference,compsoc]{IEEEtran}
\usepackage{listings}
\usepackage{acronym}
\usepackage{csvsimple}
\usepackage{adjustbox}
\usepackage{xcolor}
\usepackage{float}
\usepackage{makecell}
\usepackage{subfig}
\usepackage{tikz}
\usepackage{amsmath}
\usepackage{comment}
\usepackage{multirow}
\usepackage{cite}
\usepackage{numprint}
\usepackage{wasysym}
\graphicspath{ {./images/} }

\usepackage{listings, xcolor}

\definecolor{verylightgray}{rgb}{.97,.97,.97}

\lstdefinelanguage{Solidity}{
	keywords=[1]{anonymous, assembly, assert, balance, break, call, callcode, case, catch, class, constant, continue, constructor, contract, debugger, default, delegatecall, delete, do, else, emit, event, experimental, export, external, false, finally, for, function, gas, if, implements, import, in, indexed, instanceof, interface, internal, is, length, library, log0, log1, log2, log3, log4, memory, modifier, new, payable, pragma, private, protected, public, pure, push, require, return, returns, revert, selfdestruct, send, solidity, storage, struct, suicide, super, switch, then, this, throw, transfer, true, try, typeof, using, value, view, while, with, addmod, ecrecover, keccak256, mulmod, ripemd160, sha256, sha3}, 
	keywordstyle=[1]\color{blue}\bfseries,
	keywords=[2]{address, bool, byte, bytes, bytes1, bytes2, bytes3, bytes4, bytes5, bytes6, bytes7, bytes8, bytes9, bytes10, bytes11, bytes12, bytes13, bytes14, bytes15, bytes16, bytes17, bytes18, bytes19, bytes20, bytes21, bytes22, bytes23, bytes24, bytes25, bytes26, bytes27, bytes28, bytes29, bytes30, bytes31, bytes32, enum, int, int8, int16, int24, int32, int40, int48, int56, int64, int72, int80, int88, int96, int104, int112, int120, int128, int136, int144, int152, int160, int168, int176, int184, int192, int200, int208, int216, int224, int232, int240, int248, int256, mapping, string, uint, uint8, uint16, uint24, uint32, uint40, uint48, uint56, uint64, uint72, uint80, uint88, uint96, uint104, uint112, uint120, uint128, uint136, uint144, uint152, uint160, uint168, uint176, uint184, uint192, uint200, uint208, uint216, uint224, uint232, uint240, uint248, uint256, var, void, ether, finney, szabo, wei, days, hours, minutes, seconds, weeks, years},	
	keywordstyle=[2]\color{teal}\bfseries,
	keywords=[3]{block, blockhash, coinbase, difficulty, gaslimit, number, timestamp, msg, data, gas, sender, sig, value, now, tx, gasprice, origin},	
	keywordstyle=[3]\color{violet}\bfseries,
	identifierstyle=\color{black},
	sensitive=true,
	comment=[l]{//},
	morecomment=[s]{/*}{*/},
	commentstyle=\color{gray}\ttfamily,
	stringstyle=\color{red}\ttfamily,
	morestring=[b]',
	morestring=[b]"
}

\usepackage[figuresleft]{rotating}

\def\dowcirc{\rotatebox[origin=c]{90}{\LEFTcircle}}

\usepackage{siunitx}
\DeclareSIUnit{\nothing}{\relax}
\DeclareSIUnit{\days}{days}
\usepackage{numprint}
\npdecimalsign{.}
\npthousandsep{,}

\usepackage{url}
\usepackage{breakurl}
\usepackage{xurl}
\usepackage{hyperref}
\usepackage[flushleft]{threeparttable}
\usepackage{cleveref}

\usepackage{enumitem}

\newlist{questions}{enumerate}{2}
\setlist[questions,1]{label=RQ\arabic*.,ref=RQ\arabic*}
\setlist[questions,2]{label=(\alph*),ref=\thequestionsi(\alph*)}

\newcommand{\sourceBig}{SCD }
\newcommand{\sourceBigNoSpace}{SCD}
\newcommand{\byteBig}{BCD }
\newcommand{\byteBigNoSpace}{BCD}
\newcommand{\GTReentrancy}{RGT }
\newcommand{\GTReentrancyNoSpace}{RGT}
\newcommand{\GTAudits}{AGT }
\newcommand{\GTAuditsNoSpace}{AGT}

\begin{document}
\title{Vulnerability Scanners for Ethereum Smart Contracts: A Large-Scale Study\vspace{-0.5cm}}

{\author
	{\IEEEauthorblockN{Christoph Sendner\IEEEauthorrefmark{1},
			Lukas Petzi\IEEEauthorrefmark{1},
			Jasper Stang\IEEEauthorrefmark{1}, and
   Alexandra Dmitrienko\IEEEauthorrefmark{1}}
		\IEEEauthorblockA{\IEEEauthorrefmark{1}University of W\"urzburg, Germany}
}}

\maketitle
\vspace{-0.5cm}

\begin{abstract}
Ethereum smart contracts, which are autonomous decentralized applications on the blockchain that manage assets often exceeding millions of dollars, have become primary targets for cyberattacks. In 2023 alone, such vulnerabilities led to substantial financial losses exceeding a billion of US dollars. 
To counter these threats, various tools have been developed by academic and commercial entities to detect and mitigate vulnerabilities in smart contracts. Our study investigates the gap between the effectiveness of existing security scanners and the vulnerabilities that still persist in practice.
We compiled four distinct datasets for this analysis. The first dataset comprises 77,219 source codes extracted directly from the blockchain, while the second includes over 4 million bytecodes obtained from Ethereum Mainnet and testnets. The other two datasets consist of nearly 14,000 manually annotated smart contracts and 373 smart contracts verified through audits, providing a foundation for a rigorous ground truth analysis on bytecode and source code. 
Using the unlabeled datasets, we conducted a comprehensive quantitative evaluation of 17 vulnerability scanners, revealing considerable discrepancies in their findings. Our analysis of the ground truth datasets indicated poor performance across all the tools we tested. This study unveils the reasons for poor performance and underscores that the current state of the art for smart contract security falls short in effectively addressing open problems, highlighting that the challenge of effectively detecting vulnerabilities remains a significant and unresolved issue.
\end{abstract}

\section{Introduction} 
\label{sec:introduction}
Blockchains are digital ledgers that enable mutually untrusted parties to transact without involving a third-party intermediary such as a bank. The rise and wide usage of blockchain platforms like Bitcoin~\cite{bitcoin}, Ethereum~\cite{ethereum-whitepaper}, and Hyperledger~\cite{Hyperledger} fueled rapid growth of the blockchain ecosystem.

Smart contracts, which are self-executing computer programs, are a crucial component of blockchain-based systems. They are deployed on a blockchain and facilitate the creation of new decentralized applications (dApp)~\cite{dapp_bitcoin}, such as Decentralized Finance (DeFi)~\cite{defi_techtarget}, Non-Fungible Tokens (NFTs)~\cite{nft_investopedia}, and games~\cite{cryptokitties}. 
Smart contracts govern the use of cryptocurrency -- digital coins that are locked in a contract and can be accessed only when specific conditions are met. Developers can use them to build various services, for instance, for anonymizing money (e.g.,  TornadoCash~\cite{tornadocash}), lending money (e.g., MakerDAO~\cite{makerdao}), and pooling resources for various projects (e.g., Uniswap~\cite{uniswap}). 

Decentralized applications have gained immense popularity, with MakerDAO alone holding 7.9 billion in cryptocurrency. However, these dApps are often targeted by malicious actors who try to exploit vulnerabilities in their contracts. Unfortunately, some of these attacks are successful and can result in damages worth millions of dollars for all parties involved~\cite{dao_hack}. 

Cryptocurrency hacks have become increasingly common in recent years, with the DAO hack~\cite{dao_hack} being the first and one of the most well-known incidents. It resulted in a loss of \$60 million US dollars and led to a hard fork of Ethereum. Other examples include the Safemoon Hack, which occurred due to an access control vulnerability and enabled adversaries to exfiltrate around 8.9 million dollars~\cite{safemoon_hack}. Furthermore, the LendHub hack took place when an attacker exploited a wrong update mechanism to steal approximately 6 million dollars~\cite{lendhub_hack}. In the Deus Finance hack, an attacker exploited an access control issue to steal 13.4 million dollars~\cite{deusfinance_hack}.

Smart contract developers face challenges in dealing with vulnerabilities and bugs, as the most traditional approach of code patching is not applicable due to the immutability of the underlying blockchain. They have to resort to smart contract update mechanisms, which, as we can see from the LendHub example, can themselves have errors and lead to exploitable vulnerabilities. Alternatively and preferably, smart contracts have to be subjected to rigorous code analysis in pre-deployment phase, before the code is uploaded to the blockchain and becomes immutable.

In recent years, many vulnerability detection tools have been developed for smart contracts that can aid smart contract developers in pre-deployment security testing. These tools can be classified into four categories: Symbolic analysis \cite{maian_paper,oyente_paper,teether_paper,securify_paper}, static analysis \cite{slither_paper,smartcheck_paper}, machine learning approaches \cite{gnnscvuldetector_paper,escort,lutz2021escort}, and fuzzing \cite{confuzzius_paper,sfuzz_paper,smartian_paper}. 
Interestingly, even though some of these tools have been available for years and provided under open-source licenses, smart contracts continue to suffer from vulnerabilities that malicious actors can exploit. In 2023 alone, hackers managed to exploit vulnerabilities in smart contracts, resulting in gains exceeding one billion US dollars~\cite{hacks_2023}. 

In this work, we aim to shed light on the problem of smart contract security testing. In particular, we want to identify the cause of the problem and answer the following questions: \emph{Why, despite the existence of many effective vulnerability detection tools, does the problem of vulnerabilities in smart contracts still prevail?  Is it due to difficulties of setting up and using those tools, which raises the adaption barrier, or because they are less effective than the security research community believes? If some tools are better suitable for the detection of selected vulnerability types, can one achieve better detection by using several tools?}

To answer these questions, we want to study existing vulnerability detection tools with the goal of accessing and comparing their effectiveness. Overall, we aim to understand if the research field devoted to vulnerability detection in smart contracts is sufficiently researched or if there are remaining open problems.  

We acknowledge the efforts of previous works that have attempted already to compare various tools for vulnerability detection~\cite{tang2021vulnerabilities,rameder2022review,zhou2022state,praitheeshan2019security,sayeed2020smart,kushwaha2022systematic,durieux2020empirical,ferreira2020smartbugs,di2023smartbugs,ghaleb2020effective,ren2021empirical,kushwaha2022ethereum,dika2018security,qian2022smart}. 
However, these works either provide a survey-like comparison without performing actual performance evaluations~\cite{tang2021vulnerabilities,rameder2022review,zhou2022state,praitheeshan2019security,sayeed2020smart,kushwaha2022systematic} or evaluate a small subset of tools on a very limited dataset~\cite{durieux2020empirical,ghaleb2020effective,ren2021empirical,kushwaha2022ethereum,dika2018security,qian2022smart}. In addition, some of these works~\cite{ghaleb2020effective,ren2021empirical,dika2018security,ferreira2020smartbugs} only analyze tools that operate either on bytecode or source code and do not offer a comprehensive exploration of both. Furthermore, previous studies have not explored the potential of bundling different tools to enhance vulnerability detection. 

In our research, we seek to understand why vulnerabilities continue to be exploited despite the existence of various scanning tools. In particular, we aim to answer the above-postulated questions by means of conducting a comprehensive study of the existing tools. This study is to be performed using a range of publicly accessible resources and extensive datasets, including bytecode, source code, and ground-truth data.

\noindent\textbf{Contributions.}
We make the following contributions: 

\begin{itemize}
\item We conduct a large study of existing vulnerability detection tools by building, using, and comparing 17 smart contract vulnerability scanners from different methodology categories that utilize static analysis, symbolic execution, fuzzing, and machine learning techniques for detection. Among them, 13 tools perform source code-based analysis, while four detect vulnerabilities at the bytecode level. Three of the tools allow for the analysis of both bytecode and source code. 
Related literature considered only a subset of tools~\cite{durieux2020empirical,ren2021empirical,dika2018security,qian2022smart} or discussed only a particular methodology~\cite{ghaleb2020effective,kushwaha2022ethereum}.
Unlike previous large-scale analysis~\cite{ren2021empirical}, our study integrates an analysis of bytecode, conducts a comprehensive analysis of the employed tools, and is not limited to just reentrancy bugs but considers eight distinct vulnerabilities. Furthermore, our research is underpinned by a more extensive dataset and a broader range of tools, setting a new benchmark in the field of smart contract vulnerability analysis.

\item To conduct our study using tools that expect source code as input, we built a dataset comprising~\numprint{77219} unique and carefully de-duplicated smart contracts. 
We collected the source codes from EtherScan~\cite{etherscan} and InterPlanetary File System (IPFS)~\cite{ipfs}, where some developers upload their source code. 
This is the largest dataset of real-world source codes directly from Ethereum blockchain available to date. 
We plan to make it available to the community for further research. 

\item To study tools operating on bytecode of smart contracts, we built an unprecedentedly large dataset of~\numprint{4062844} bytecodes after de-duplication of initially collected~\numprint{26740370} bytecodes. For this dataset, we tried to get as many Ethereum-compatible smart contracts as one could find. In particular, we collected all smart contracts from Ethereum network and extended it with smart contracts collected from four test networks: Goerli \cite{testnet-goerli}, Rinkeby \cite{testnet-rinkeby}, Ropsten \cite{testnet-ropsten}, and Kovan \cite{testnet-kovan}. We plan to make this dataset available as well.

\item We performed a comprehensive study of 13 source code-based tools (Slither~\cite{slither_paper}, SmartCheck~\cite{smartcheck_paper}, Maian~\cite{maian_paper}, Oyente~\cite{oyente_paper}, Artemis~\cite{artemis_paper}, Osiris~\cite{osiris_paper}, Securify2~\cite{securify_paper}, Mythril~\cite{mythril_paper}, TeEther~\cite{teether_paper}, ConFuzzius~\cite{confuzzius_paper}, Smartian~\cite{smartian_paper}, sFuzz~\cite{sfuzz_paper}, GNNSCVulDetector~\cite{gnnscvuldetector_paper}) 
and attempted to detect 8 vulnerability types in our source code-based dataset. The goal of this study was to establish if the outcomes of these tools are consistent with each other, and if one could potentially use a strategy of using several detection tools to enhance detection performance. The outcomes of this study are non-trivial -- we observed significant discrepancies in detection outcomes. For instance, we reveal that the tools do not agree on a single sample to have the reentrancy vulnerability. 
This shows that the idea of combining several tools won't be very useful in practice.

\item We additionally studied four vulnerability detection tools that operate on the byte-code level: Vandal~\cite{vandal_paper}, Maian~\cite{maian_paper}, Oyente~\cite{oyente_paper}, Mythril~\cite{mythril_paper}. 
Their performance was evaluated using 5 vulnerability types. The results of this evaluation similarly display lack of consensus among the tools, to the extent that questions their ability for accurate detection. 

\item Significant discrepancies in detection outcomes of both source code and bytecode-based tools indicate that at least some of them do not provide good detection performance. To verify this preposition, we built two additional datasets labeled with ground-truth labels. Our team manually labeled one ground truth dataset for the reentrancy bug, consisting of 13,773 unique smart contracts. The other dataset comprises 373 unique smart contracts collected from publicly available repositories and includes eight vulnerability types. Vulnerabilities in these contracts were confirmed by security audits conducted by security firms. 

\item Equipped with the two ground truth datasets, we evaluate the performance of both source code and bytecode-based tools involved in our study. Our results highlight the overall poor performance of all tools under our evaluation, with F1-scores ranging from 0\% to a maximum of 73\%. This is primarily attributed to the high incidence of false positives and negatives reported by the tools.

\item The results of our analysis reveal the reasons why the performance of the tools under evaluation is often below expectations and derive other valuable insights. We provide insights into these reasons. For instance, we observe that varying compiler versions can lead to significant changes that obstruct the effectiveness of vulnerability scanners in their analysis. Additionally, the evolving coding practices of developers could also affect the performance of these scanners. Our findings aim to assist future research in developing more efficient tools for detecting vulnerabilities. 

\end{itemize}

Overall, our study demonstrates that the current state of the art in the area of smart contract security has significant room for improvement and the problem of vulnerability detection remains an open and challenging problem.

\vspace{0.2cm}
\noindent \textbf{Outline.} The rest of this paper is structured as follows: We offer in~\Cref{sec:background} a brief overview of smart contract vulnerabilities and the tools we utilized in this study to detect them. In~\Cref{sec:datasets}, we describe our datasets and the methodology employed in their construction. The analysis of the different tools is presented in~\Cref{sec:analysis}. An additional discussion around security analysis of smart contracts is provided in~\Cref{sec:discussion}. We examine related work in~\Cref{sec:related_work}. We conclude the paper in~\Cref{sec:conclusion}.
\section{Background} 
\label{sec:background}
This section provides the background information for smart contracts, their vulnerabilities, and the vulnerability scanners. We will especially focus on those vulnerabilities and scanners that will be further explored in this study.

\subsection{Smart Contracts}
\label{subsec:contracts}
A smart contract refers to a software program that operates within a blockchain environment. 
In this study, we focus on smart contracts designed for the Ethereum blockchain. These contracts are typically written in Solidity~\cite{solidity}, compiled, and executed within the Ethereum Virtual Machine (EVM). 
The EVM functions as a stack-based machine with a word size of 256 bits and a stack size of 1024, utilizing a word-addressable memory model~\cite{ethereum-yellowpaper}.

To deploy a smart contract, the compiled EVM bytecode is uploaded to the blockchain through a transaction. 
Interacting with the smart contract is also achieved by sending transactions to the contract's address. 
It is important to note that executing smart contracts incurs a cost in the form of Gas, which is essentially a fee that aims to cover the costs of execution (e.g., electricity costs).
Certain contracts incorporate an IPFS link within their bytecode by encoding essential metadata. This link grants access to retrieve valuable information, including the Application Binary Interface (ABI), source code, and additional metadata.

\subsection{Smart Contract Vulnerabilities}
\label{subsec:vulns}

Similar to any other software, smart contracts are susceptible to bugs and vulnerabilities. 
Given that smart contracts are directly associated with cryptocurrencies, the potential financial losses resulting from undiscovered vulnerabilities can be significant. 
Numerous vulnerabilities exist in the realm of Solidity smart contracts and are listed in the Smart Contract Weakness Classification Registry (SWC)~\cite{swc}. 
Overall, one can classify these vulnerabilities into three categories: Software Errors, Runtime Bugs, and Blockchain Characteristics.

\vspace{0.2cm}\noindent\textbf{Software Errors}
In the field of Solidity smart contracts, software errors often result in bugs, with arithmetic issues (SWC-101~\cite{swc}) such as integer overflows and underflows being particularly prevalent. Notably, these vulnerabilities have been partly addressed through a compiler update in Solidity. However, for comprehensive legacy detection, these issues remain a focus of this study.

Additionally, various other critical bugs warrant attention. The 'suicide' vulnerability (SWC-106~\cite{swc}), for example, can potentially allow an attacker to destroy a contract if its functions lack proper safeguards. Equally concerning is the assert violation~\cite{assert_violation}, which arises when developers inadvertently leave a failing assert statement in live code. Also of significance is the misuse of txOrigin (SWC-115~\cite{swc}), which, if exploited in place of msg.sender, could allow attackers to manipulate a global variable to their advantage.

Another notable vulnerability involves the use of block timestamps as a proxy for time, which can be exploited by attackers aware of this time dependency (SWC-116~\cite{swc}). Lastly, the issue of 'Locked Ether' arises from the absence of a mechanism to withdraw Ether from the contract, leading to potential fund loss if not addressed pre-deployment.

\vspace{0.2cm}
\noindent\textbf{Runtime Bugs}
Smart contracts also face a range of bugs associated with their execution runtime. These vulnerabilities arise due to specific characteristics of how smart contracts are executed. For instance, an example vulnerability is reentrancy (SWC-107~\cite{swc}), where an attacker can exploit the ability to reenter a function during runtime, even before the initial execution is completed. Another example is the legacy callstack depth vulnerability~\cite{callstack_depth}, where the stack can become exhausted. Additionally, there are "greedy contracts"~\cite{maian_paper}, which only accept Ether without providing a means for later extraction.
Another critical vulnerability is associated with the use of DelegateCall (SWC-112~\cite{swc}). This feature allows a contract to execute another contract's code within its own context. It's essential to emphasize that the external contract's code must be thoroughly vetted for security, as any compromise in its integrity could give an attacker complete control over the caller's funds.

\vspace{0.2cm}
\noindent\textbf{Blockchain Characteristics}
The blockchain infrastructure itself can give rise to vulnerabilities within smart contracts. An instance of such vulnerability is money concurrency or Transaction Ordering Dependency (ToD) (SWC-114~\cite{swc}), which can result in financial losses when the code depends on the sequence of transactions, such as determining the first correct answer submitted~\cite{tod_example}. Another concern emerges when developers utilize blockchain's block values for time-related purposes. However, these values can be influenced by miners and should not be relied upon. Additionally, creating randomness within smart contracts becomes challenging due to their deterministic execution nature, potentially leading to vulnerabilities.

\subsection{Vulnerability Scanners} 
\label{sec:tools}
In the following, we categorize each vulnerability scanner into one of four analysis approaches: Static Analysis, Symbolic Execution, Fuzzing, and Machine Learning.

\vspace{0.2cm}\noindent\textbf{Static Analysis} \label{subsec:static_analysis}
Static Analysis involves examining source code or compiled output without execution. This allows tools to detect bugs and security issues without requiring an execution environment or risking running vulnerable or malicious code.

\emph{Slither}\cite{slither_paper} is a static analysis framework for finding vulnerabilities in Solidity source files. It supports the detection of over 80 vulnerability classes and can be extended. Slither supports the detection of, for example, suicidal contracts, reentrancy, and block data dependency.

\emph{Vandal}\cite{vandal_paper} is a security analysis framework designed for EVM bytecodes. It operates by converting the bytecode into semantic logic relations, which are subsequently analyzed against specified vulnerabilities using a declarative language. The tool detects a range of vulnerabilities, including Unchecked Send, Reentrancy, and Selfdestruct.

\emph{SmartCheck}\cite{smartcheck_paper} conducts thorough lexical and syntactical analyses on source code to identify vulnerabilities. It employs text parsing techniques to convert Solidity source code into an XML parse tree, which is subsequently examined for vulnerability patterns. It is crucial to emphasize that as of 2020, SmartCheck has been deprecated and is no longer actively maintained.

\emph{EtherTrust}\cite{ethertrust_paper} classifies bytecodes by defining an abstract EVM semantic representation and rules for detecting reentrancy vulnerabilities. It abstracts EVM bytecode and employs static reachability analysis with Horn clauses for vulnerability detection.

\vspace{0.2cm}
\noindent\textbf{Symbolic Execution}
\label{subsec:symbolic_execution}
Differing from Static Analysis, Symbolic Execution encompasses the execution of the specific source code or compiled code, including dynamic analysis. Given the overlapping nature of symbolic execution and dynamic analysis, these tools are categorized together under this unified category.

\emph{Maian}\cite{maian_paper} is a symbolic analysis tool for EVM bytecodes and Solidity source codes. It detects three types of vulnerabilities: Prodigal, Suicidal, and Greedy contracts.

\emph{Oyente}\cite{oyente_paper} is a symbolic execution tool created for detecting possible security flaws in the source code and bytecode of Ethereum smart contracts. Like SmartCheck, Oyente is also outdated and no longer actively maintained.

\emph{Artemis}\cite{artemis_paper} and \emph{Osiris}\cite{osiris_paper} are extensions of Oyente that aim to detect additional vulnerabilities.

\emph{Securify2}\cite{securifyv2} is an analysis tool based on Securify \cite{securify_paper} that was deprecated since 2019. Similar to the initial version, Securify2 performs in-depth analysis of EVM bytecode, examining it against a predefined set of security patterns. These patterns consist of both compliance and violation patterns, which encompass the necessary conditions for ensuring the smart contract's adherence to specific security requirements or, conversely, flagging instances where predefined security properties are violated.

\emph{EthBMC}\cite{ethbmc_paper} is a bounded model checker analysis tool that utilizes symbolic execution to examine smart contract EVM bytecodes against predefined models. It specializes in detecting suicidal contracts and contracts that permit arbitrary extraction of funds.

\emph{Mythril}\cite{mythril_paper} is a EVM bytecode analysis tool. The tool combines symbolic execution, SMT solving, and taint analysis techniques to detect a wide range of vulnerabilities, such as suicidal contracts or contracts that rely on weak sources of randomness.

\emph{TeEther}\cite{teether_paper} does not primarily focus on general vulnerability detection but rather the creation of practical exploits. To achieve this, it specifically targets four low-level instructions closely tied to value transfer: (1) Call, (2) Selfdestruct, (3) Callcode, and (4) Delegatecall. By imposing path constraints and evaluating their satisfiability, the tool examines each execution path to determine if an undesired value transfer could potentially occur.

\vspace{0.2cm} \noindent\textbf{Fuzzing}
\label{subsec:fuzzing}
Although fuzzing can be categorized as a form of runtime analysis, we distinguish it from other techniques based on the selection of input data. Fuzzing involves the guided but random selection of input data, setting it apart from other methods in the Symbolic Execution category.

\emph{ILF}\cite{ilf_paper} is a fuzzing tool that classifies solidity source code files. The fuzzer is trained from data generated by a symbolic execution engine. The trained fuzzing model can be used to detect vulnerabilities such as Suicidal contracts, Leaking vulnerability or Block dependency.

\emph{ConFuzzius}\cite{confuzzius_paper} is an evolutionary fuzzer that applies constraint resolving and data dependency analysis to detect smart contract vulnerabilities. The fuzzer detects 10 vulnerabilities, including Suicidal, Reentrancy, and ToD.

\emph{Smartian}\cite{smartian_paper} is a mutation-based fuzzer that integrates static and dynamic analyses to drive input mutation. By leveraging dynamic dataflow analysis, Smartian dynamically guides the fuzzing engine and incorporates bug oracles into its testing process.

\emph{sFuzz}\cite{sfuzz_paper} is a framework developed based on AFL~\cite{afl}, a renowned fuzzer primarily used for C/C++ programs. It extends the capabilities of AFL to support EVM bytecode by introducing adaptations in multiple areas. Notably, sFuzz introduces a novel metric to target hard-to-cover branches specifically. 

\vspace{0.2cm}
\noindent\textbf{Machine Learning}
\label{subsec:machine_learning}
Machine Learning has been widely utilized in various domains to identify vulnerabilities, and the realm of smart contracts is no exception. In this context, we introduce two available tools that leverage Graph Neural Networks (GNNs) and Recurrent Neural Networks (RNNs), respectively, for vulnerability detection purposes.

\emph{GNNSCVulDetector}\cite{gnn_paper} is designed to detect vulnerabilities in smart contracts by utilizing a GNN. It analyzes the syntactic and semantic structures of the smart contract, which are represented as a graph. This graph is processed using a degree-free graph convolutional neural network for classification. The tool identifies three vulnerability classes: Reentrancy, Timestamp Dependency, and Infinite Loop. While the source code for the generation of the datasets and training the model is publicly available, execution is restricted. With minor bug fixes the dataset generation and training was successful for the timestamp vulnerability but not for Reentrancy and Infinite Loop vulnerabilities.
\section{Datasets} 
\label{sec:datasets}
Ethereum contracts are stored across various blockchains, primarily encompassing the Ethereum Mainnet (i.e., the primary production blockchain)
and the testnets, namely Goerli \cite{testnet-goerli}, Rinkeby \cite{testnet-rinkeby}, Ropsten \cite{testnet-ropsten}, and Kovan \cite{testnet-kovan}. We successfully generated four distinct datasets from those blockchains. The first dataset is the Source Code Dataset (\sourceBigNoSpace), which comprises a collection of smart contract source codes. The second dataset is the Bytecode Dataset (\byteBigNoSpace), consisting of a hexadecimal representation of compiled bytecodes of the smart contracts. The third dataset is the Reentrancy Ground Truth (\GTReentrancyNoSpace), specifically focused on identifying and labeling source code operations that are vulnerable to reentrancy. Lastly, the fourth dataset is the Audits Ground Truth dataset (\GTAuditsNoSpace), which includes labeled data obtained from security audits conducted on smart contracts. 

\subsection{Source Code Dataset}
\label{sub:source_code_data}
The process of assembling our source code dataset involved two distinct stages. In the initial phase, we used Google BigQuery~\cite{google-bigquery} to gather addresses of smart contracts. The second phase entailed downloading the source code, which was accessible either through EtherScan~\cite{etherscan} or IPFS~\cite{ipfs}.

EtherScan~\cite{etherscan}, a prominent blockchain explorer and analytics tool for the Ethereum network, offers a detailed set of functionalities for examining and extracting information about Ethereum transactions, addresses, smart contracts, and overall network dynamics. One of its key features is providing access to the source code of specific contracts on the Mainnet, enabling a deeper understanding of the contracts' operations and verifying their functionalities. We utilized the EtherScan API to systematically download all accessible source codes from the Mainnet blockchain. 

Regarding IPFS, our approach started with downloading the contract bytecodes to check for embedded metadata. This metadata, located at the end of the bytecode, was then decoded using a CBOR decoder. From the decoded information, we extracted the IPFS link, which led us to the original contract metadata containing the source code.

After implementing a deduplication process by hashing the smart contracts and comparing these hashes, we successfully compiled a dataset of~\numprint{77219} unique source codes, all written in the Solidity language.

\subsection{Bytecode Dataset}
\label{sub:bytecode_data}
We retrieve the bytecode from Ethereum contracts on different blockchains. 
To synchronize with the Ethereum blockchain, we employed two clients: Erigon~\cite{erigon-client} for the Goerli, Ropsten, and Rinkeby testnets, chosen for its speed and lower disk space usage, and Geth~\cite{goethereum-client} for the Kovan network, as it is not yet supported by Erigon. Furthermore, Ethereum Mainnet contracts were sourced from an open dataset available through Google's BigQuery service~\cite{google-bigquery}.

We proceeded by retrieving the contract addresses from each blockchain and utilizing the Python Web3 API library \cite{web3-docs} to extract the EVM bytecode from the downloaded blockchains. These extracted bytecodes were then stored in a MySQL database. Some contracts obtained this way were found to be empty. There are a few possible reasons for this. It could be due to the Ethereum node not being fully synchronized with the network, resulting in an unavailable bytecode. Alternatively, it could be because empty contracts were deployed on the blockchain or the contract had been self-destructed. Overall, we successfully extracted \numprint{26740370} non-empty bytecodes from the five networks. Specifically, the Ethereum Mainnet contributed \numprint{22789100} bytecodes, Ropsten provided \numprint{1831168}, Rinkeby contributed \numprint{1382338}, while Kovan and Goerli supplied \numprint{635766} and \numprint{101998} bytecodes, respectively. After deduplication, we can use~\numprint{4062844} unique bytecodes for our analysis. 

\subsection{Reentrancy Ground Truth}
\label{sub:reentrancy_data}
This dataset is centered explicitly on reentrancy vulnerabilities, selected due to their frequent detection by smart contract vulnerability scanners. This focus allows us to compare across multiple tools. Given the intensive nature of manually labeling numerous contracts, our study is concentrated exclusively on this particular vulnerability. Consequently, we have developed a specialized source code dataset enriched with detailed annotations to deepen our investigation into reentrancy vulnerabilities.
As a foundation, we utilized the SmartBugs Wild Dataset \cite{dataset_smartbugswild}, which encompasses \numprint{47398} source codes from various smart contracts. 
We began by eliminating duplicates from the dataset and then proceeded to incorporate the annotations specific to the reentrancy vulnerability. 
This dataset allows for a more detailed analysis of how effectively reentrancy vulnerabilities are detected in smart contract source codes and bytecodes downloaded directly from the blockchain.

\vspace{0.2cm}
\noindent\textbf{Preprocessing} During the preprocessing step, our focus was on removing duplicates from the dataset. These duplicates often arise when source code is copied and subsequently subjected to minor modifications, such as comment adjustments, variable or function renaming, or changes to variable values. To identify such contracts, we utilized the Solidity compiler to generate an Abstract Syntax Tree (AST) from the source code file. This AST representation allowed us to eliminate comments and whitespace characters while also removing intermediate values, variable names, and function names. This process facilitated effective comparison between Solidity files. By assessing the similarity of the AST trees by comparing their hashes, we identified contracts that displayed little resemblance to others, resulting in a refined dataset comprising of \numprint{22237} source code files.

\vspace{0.2cm}
\noindent\textbf{Annotation} Subsequently, we proceeded to annotate the deduplicated dataset. The original SmartBugs Wild dataset only provided annotations for the reentrancy vulnerability for the call subtype. To achieve a more comprehensive and detailed annotation, we expanded our analysis to include all three subtypes: call, send, and transfer. 
The 'call' subtype is particularly critical, as its function lacks intrinsic gas limitations. 
On the other hand, 'send' and 'transfer' subtypes are designed to use only a specified amount of gas for the contract call, as defined in the Ethereum Yellow Paper~\cite{ethereum-yellowpaper}. 
However, this amount may vary with updates to the Ethereum network.
It is important to note that contracts lacking these subtypes were considered non-vulnerable.

For each contract, we manually inspected the source code containing the three subtype functions. Our assessment focused on determining whether a state change occurred after the transfer of funds and whether a reentrancy occurred. Upon meeting these conditions, we annotated the respective contract as vulnerable to reentrancy attacks. As a result, our final reentrancy dataset comprises \numprint{13773} smart contracts. 

\subsection{Audits Ground Truth}
\label{sub:audits_data}
Obtaining a ground truth dataset for evaluating the detection performance of various tools can be challenging and time-consuming. It requires significant investment in correctly labeling the data, often involving the expertise of human auditors. This process, known as security audits, ensures that the presence of vulnerabilities in smart contract source code is accurately identified. To acquire a reliable ground truth dataset for tools that operate at the level of the source code, we utilized publicly available audit repositories from reputable sources, including Quantstamp~\cite{quantstamp}, OpenZeppelin~\cite{openzeppelinaudits}, Trail of Bits~\cite{trailofbitsaudits}, ConsenSys~\cite{consensysaudits}, and CertiK~\cite{certikaudits}. 
The resulting dataset consists of \numprint{373} smart contract source codes that have been labeled by security auditors into the vulnerability categories. 
This carefully labeled ground truth dataset provides a robust benchmark for evaluating the performance of different tools in vulnerability detection. Further, we compiled the available source codes to allow a bytecode-based analysis.
\section{Study} 
\label{sec:analysis}
In this section, we detail our study conducted with 13 source code-based and four bytecode-based tools, using the four datasets described in \Cref{sec:datasets}. The \sourceBig and \byteBig datasets enable quantitative evaluation, while the \GTReentrancy and \GTAudits datasets facilitate qualitative analysis of the tools.

\subsection{Methodology}
\label{sub:methodology}
In this section, we will discuss our methodology for the analysis of source codes, bytecode, and describe our visualization method. 

We use the vulnerability scanners as-is and do not optimize them per smart contract. For instance, if the tool supports only a limited range of compiler versions, we don't attempt to enhance it to other versions. We also consider vulnerability scanners that time out on a smart contract as non-vulnerable since the outcome is the same for a developer -- no vulnerability is found or reported. For a detailed evaluation of scanning robustness, we refer the reader to \Cref{sub:robustness_eval}, where we analyze the completion rate of the different scanners.

\vspace{0.2cm}\noindent\textbf{Source code analysis:} We employed 13 tools to identify over \numprint{200} vulnerability types. For comparability and simplicity, our focus was on eight types detectable by at least three scanners. Results for other types are omitted due to space limitations.

\vspace{0.2cm}\noindent\textbf{Bytecode analysis} We analyzed the bytecode-based datasets using four tools. Since the tools' density is lower than that of those operating on source codes, we provided our analysis of tools based on four vulnerability types present across the tools.

\vspace{0.2cm}\noindent\textbf{Visualization} We opted for Upset plots that proved to be superior to Venn diagrams for visualizing complex intersections in datasets, as they provide a clearer and more scalable representation of the relationships between multiple sets, especially when dealing with large numbers of sets where Venn diagrams become cluttered and less interpretable.

Each plot comprises two main sections: On the left side, we display the total number of vulnerable samples found by each tool. On the top, we show the total number of samples that overlap among the tools. If there is a single dot, it signifies that these samples do not overlap with any other tool. Conversely, when a column contains multiple dots, it signifies that the respective tools agree in their analysis of these specific samples.

\vspace{0.2cm}\noindent\textbf{Vulnerabilities}
We focus on eight types of vulnerability throughout this paper: Suicide, Reentrancy, Transaction Order Dependency (ToD), Arithmetic Bugs, Usage of txOrigin, Time Dependency, Locked Ether, and DelegateCall.

\vspace{0.2cm}\noindent\textbf{Test Environment} All tests were conducted in our High-Performance Cluster, in which each node comprises two Intel® Xeon Gold 6134 Processors (8c/16t), with 384 GB DDR4 memory and BeeGFS~\cite{beegfs} for storage. We use Docker~\cite{docker} to parallelize the use of different tools and, thus, maximize the usage of available resources in the cluster. To get the tools running and thoroughly test all tools on the datasets, we invested significant manual effort over twelve person-months.

\subsection{Quantitative Analysis on \sourceBig Dataset}
\label{sub:source_eval}

As per our analysis, we have evaluated all~\numprint{77219} smart contracts available in \sourceBig using 13 source code-based tools, listed in \Cref{tab:overview_tools_source}. We utilized them to detect eight types of vulnerabilities, namely: \textit{Suicide}, \textit{Reentrancy}, \textit{Transaction Order Dependency (ToD)}, \textit{Arithmetic Bugs}, \textit{Usage of txOrigin}, \textit{Time Dependency}, \textit{Locked Ether}, and \textit{DelegateCall}.
During our analysis, we found that some tools provide a more detailed analysis of the reentrancy vulnerability type by reporting vulnerability sub-types. For example, the reentrancy vulnerability type can be divided into \textit{bad}, \textit{No Eth}, or \textit{Benign}, as per the analysis by Slither and Securify2. Additionally, vulnerabilities like \textit{Arithmetic Bugs} have been grouped based on the type of bugs, such as \textit{Integer Overflow/Underflow} and generic \textit{Arithmetic Bugs}. 

\begin{table*}[ht!]
	\centering
	\scalebox{0.8}{
\begin{tabular}{|c|c|c|c|c|c|c|c|c|c|c|c|c|c|}
\hline
\multirow{2}{*}{\thead{Metrics}} & \multicolumn{2}{l|}{ \thead{Static Analysis}} & \multicolumn{7}{l|}{\thead{Symbolic Execution}} & \multicolumn{3}{l|}{\thead{Fuzzing}} & \thead{Machine Learning} \\ \cline{2-14} 
                    & Slither     & SmartCheck   &  Maian   & Oyente   & Artemis   &  Osiris   & Securify2  & Mythril  & TeEther  & ConFuzzius & Smartian & sFuzz    & GNNSCVulDetector \\ \hline
Suicidal           &   \CIRCLE    &  \Circle\    & \CIRCLE  & \Circle\ & \Circle\  & \Circle\  & \CIRCLE    & \CIRCLE  & \CIRCLE  & \CIRCLE    & \CIRCLE  & \Circle\ & \Circle\ \\ 
Reentrancy         &   \dowcirc   &  \Circle\    & \Circle\ & \CIRCLE  & \CIRCLE   & \CIRCLE   & \dowcirc   & \CIRCLE  & \Circle\ & \CIRCLE    & \CIRCLE  & \CIRCLE  & \Circle\ \\ 
ToD                &   \Circle\   &  \Circle\    & \Circle\ & \CIRCLE  & \CIRCLE   & \CIRCLE   & \dowcirc   & \Circle\ & \Circle\ & \CIRCLE\   & \Circle\ & \Circle\ & \Circle\ \\ 
Arithmetic Bug     &   \Circle\   &  \Circle\    & \Circle\ & \dowcirc & \dowcirc\ & \dowcirc\ & \Circle\   & \CIRCLE  & \Circle\ & \dowcirc\  & \CIRCLE\ & \Circle\ & \Circle\ \\ 
TxOrigin           &   \CIRCLE    &  \CIRCLE     & \Circle\ & \Circle\ & \CIRCLE   & \Circle\  & \CIRCLE    & \CIRCLE  & \Circle\ & \Circle\   & \CIRCLE  & \Circle\ & \Circle\ \\ 
Timestamp          &   \CIRCLE    &  \Circle\    & \Circle\ & \CIRCLE  & \CIRCLE   & \CIRCLE   & \CIRCLE    & \CIRCLE  & \Circle\ & \CIRCLE    & \CIRCLE  & \CIRCLE  & \CIRCLE  \\ 
Lock               &   \CIRCLE    &  \CIRCLE     & \CIRCLE  & \Circle\ & \Circle\  & \Circle\  & \CIRCLE    & \Circle\ & \Circle\ & \CIRCLE    & \Circle\ & \CIRCLE  & \Circle\ \\ 
DelegateCall       &   \dowcirc   &  \Circle\    & \Circle\ & \Circle\ & \CIRCLE   & \Circle\  & \CIRCLE    & \CIRCLE  & \CIRCLE  & \CIRCLE    & \Circle\ & \CIRCLE  & \Circle\ \\ \hline
\end{tabular}
}
   \caption{Overview of vulnerability scanners with detectable vulnerability types on source code.}   \label{tab:overview_tools_source}
\end{table*}

\begin{figure*}[ht!]
\centering
\includegraphics[width=2.0\columnwidth]{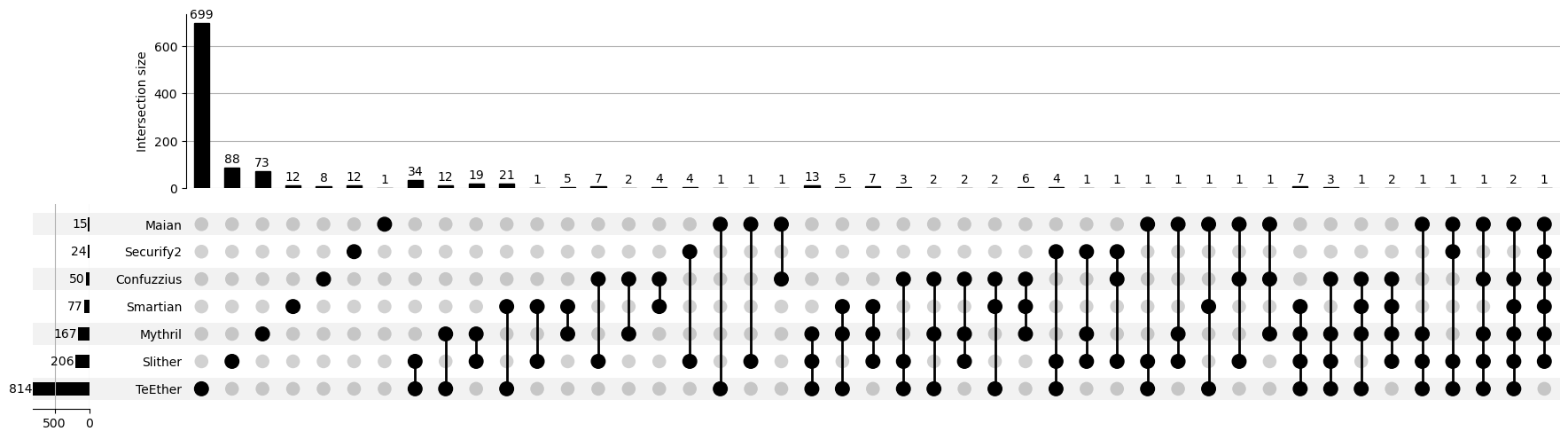}
\vspace{-1.4em}
\caption{Overlap of source code-based tools detecting the \textit{Suicide} vulnerability.
}\label{fig:source_suicide}
\vspace{0.5em}
\end{figure*}

\vspace{0.2cm} \noindent\textbf{Suicide}
We utilized seven distinct tools to identify occurrences of suicidal contracts in the source code. These tools are Slither, Mythril, Smartian, Confuzzius, Securify2, Maian, and TeEther. In \Cref{fig:source_suicide}, we have visually presented the results obtained from running these tools. 
The figure clearly illustrates that none of the contracts were identified as vulnerable by all seven tools. The highest level of agreement was six out of seven tools, and this only occurred for three smart contracts, as indicated in the figure's last two rows. Summing up the columns that contain at least three dots reveals that a mere 70 contracts were marked as vulnerable by three or more tools.

The overlap between the scanners is minimal, with four tools independently detecting the suicide vulnerability in over \SI{50}{\percent} of all tool-flagged contracts. 
Slither and TeEther have the most overlap with other tools in this analysis. However, even in these cases, the overlap with other tools is still limited. 

\begin{figure*}
\centering
\includegraphics[width=2.0\columnwidth]{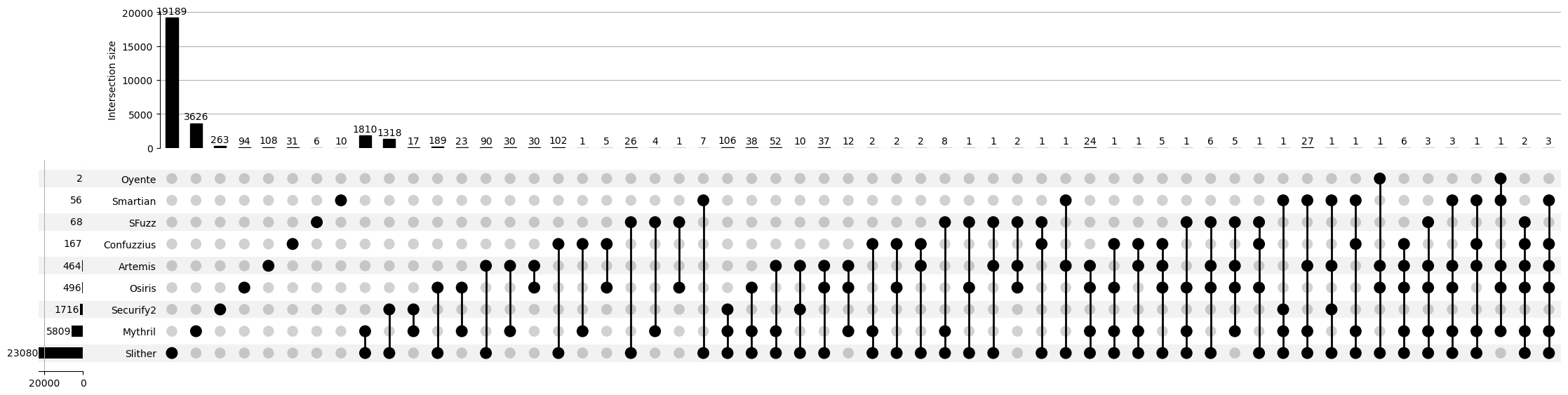}
\vspace{-1.4em}
\caption{Overlap of source code-based tools detecting the \textit{Reentrancy} vulnerability.
}\label{fig:source_reentrancy}
\vspace{0.5em}
\end{figure*}

\noindent\textbf{Reentrancy}
Our study shows that there is minimal overlap between different tools when it comes to detecting the reentrancy vulnerability in smart contracts, as illustrated in \Cref{fig:source_reentrancy}. We could utilize nine different tools to detect this vulnerability, thanks to the popularity of this vulnerability type among various scanners. Although Slither stands out from the rest with an extremely high positive rate, including the positives of other tools, the most significant finding is that no single contract is marked as vulnerable when all the tools' results are combined. Out of the total of~\numprint{77219} smart contracts checked, only \numprint{106} were identified as vulnerable by three tools at most. This highlights the need for a comparative study of various tools to comprehensively and accurately assess this smart contract vulnerability.

\begin{figure}[ht!]
\centering
\includegraphics[width=0.9\columnwidth]{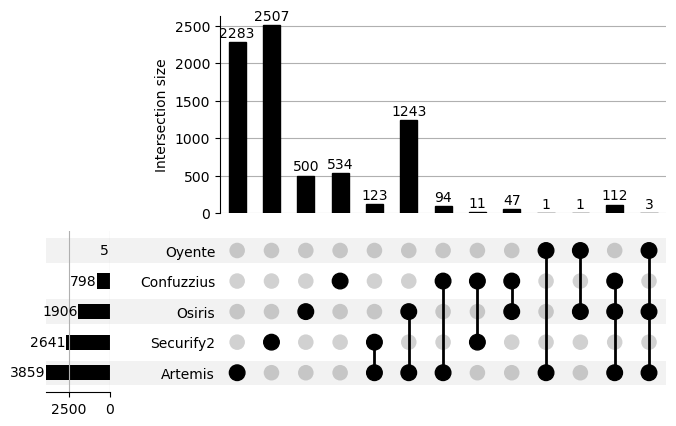}
\caption{Overlap of source code-based tools detecting the \textit{ToD} vulnerability.
}\label{fig:source_tod}
\end{figure}

\noindent\textbf{ToD} 
In our evaluation of the ToD vulnerability, as shown in \Cref{fig:source_tod}, none of the smart contracts were identified as vulnerable by all five tested tools. 
Additionally, not even four out of five tools agreed on a single contract. 
However, there was a significant overlap between Osiris and Artemis, which is not surprising since they are both based on the same underlying tool, Oyente. 
Oyente also overlaps with Artemis and Osiris, although it provides the fewest positive detections.

\begin{figure*}
\centering
\includegraphics[width=2.0\columnwidth]{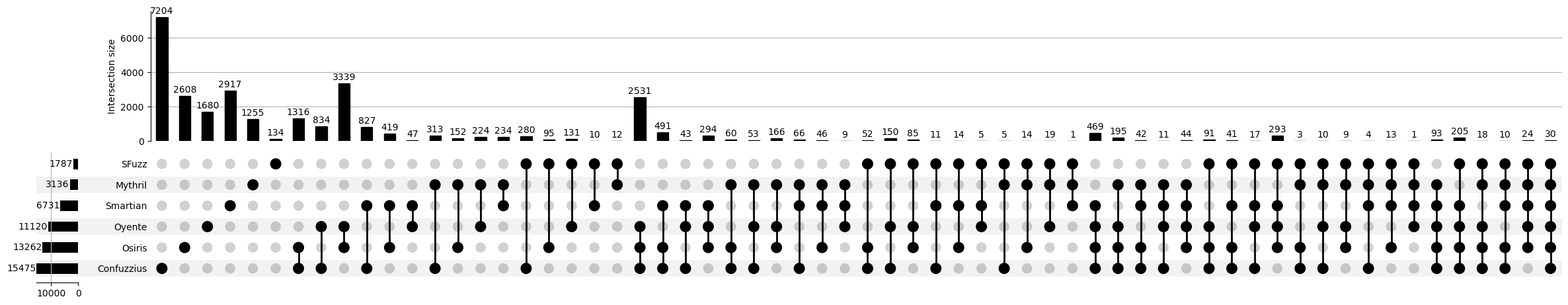}
\vspace{-1.4em}
\caption{Overlap of source code-based tools detecting the \textit{Arithmetic} vulnerability.
}\label{fig:source_ari}
\vspace{0.5em}
\end{figure*}

\noindent\textbf{Arithmetic Bugs}
The evaluation results for reported arithmetic bugs are presented in \Cref{fig:source_ari}, which displays the test results for six underlying tools. Remarkably, Artemis is notably absent from the figure as it did not detect a single bug.
Unlike previously observed trends, the tools agreed on \numprint{30} cases. 
It is noteworthy that Oyente, Osiris, and Confuzzius showed a significant overlap and agreement on an arithmetic bug in \numprint{2622} samples. 
While the similarity of Osiris and Oyente can be explained by the fact that Osiris is based on Oyente, the reasons for the overlap between Confuzzius and Oyente are not that apparent. Generally, the overlap between the tools is more pronounced than with the other vulnerabilities.
We attribute this result to the simplistic nature of this vulnerability type. But we also note that this positive result is undermined by the fact integer over- and underflow vulnerability is mitigated directly by the compiler. 

\begin{figure}[ht!]
\centering
\includegraphics[width=1.0\columnwidth]{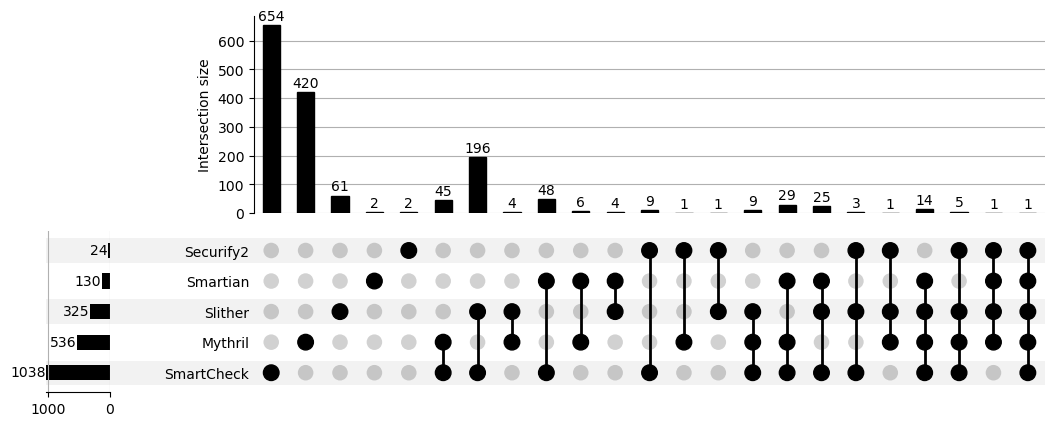}
\vspace{-1.4em}
\caption{Overlap of source code-based tools detecting the \textit{txOrigin} vulnerability.
}\label{fig:source_txo}
\vspace{0.5em}
\end{figure}

\noindent\textbf{txOrigin}
Based on our evaluation results, \Cref{fig:source_txo} indicates that all five analyzed tools agree on only one sample to have the vulnerability. 
We observed that Slither overlaps with SmartCheck for most of its detected samples, which is quite interesting. 
Other than that, the tools mostly disagree, similar to the other vulnerabilities.

\noindent\textbf{Time Dependency}
We use ten security scanners to identify the Time Dependency vulnerability.
The \Cref{fig:source_time} from the~\Cref{sec:appendix} shows our evaluation results where not a single contract was identified as vulnerable by all tools.
Eight tools agree on the existence of the vulnerability in seven samples out of our dataset.
Similar to the reentrancy vulnerability, Slither identifies the most potentially vulnerable samples.
Therefore, most of the other tools have a significant overlap with Slither.
However, the majority of positive samples are disjoint. 

\begin{figure}[ht!]
\centering
\includegraphics[width=1.0\columnwidth]{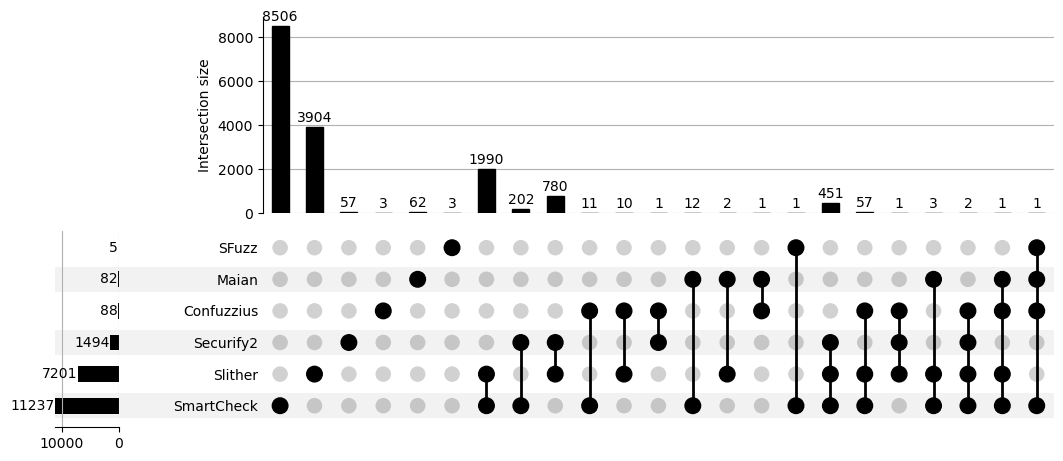}
\vspace{-1.4em}
\caption{Overlap of tools detecting the \textit{Locked Ether} vulnerability.
}\label{fig:source_lock}
\vspace{0.5em}
\end{figure}

\noindent\textbf{Locked Ether} 
According to the evaluation results presented in \Cref{fig:source_lock}, none of the six tools agree on a single sample. SmartCheck flags the most contracts as vulnerable and has a significant overlap with Slither. Securify2 has most of its flagged contracts overlapping with Slither and SmartCheck. Maian, on the other hand,  only overlaps with the other tools on \numprint{20} samples. 

\begin{figure}[ht!]
\centering
\includegraphics[width=1.0\columnwidth]{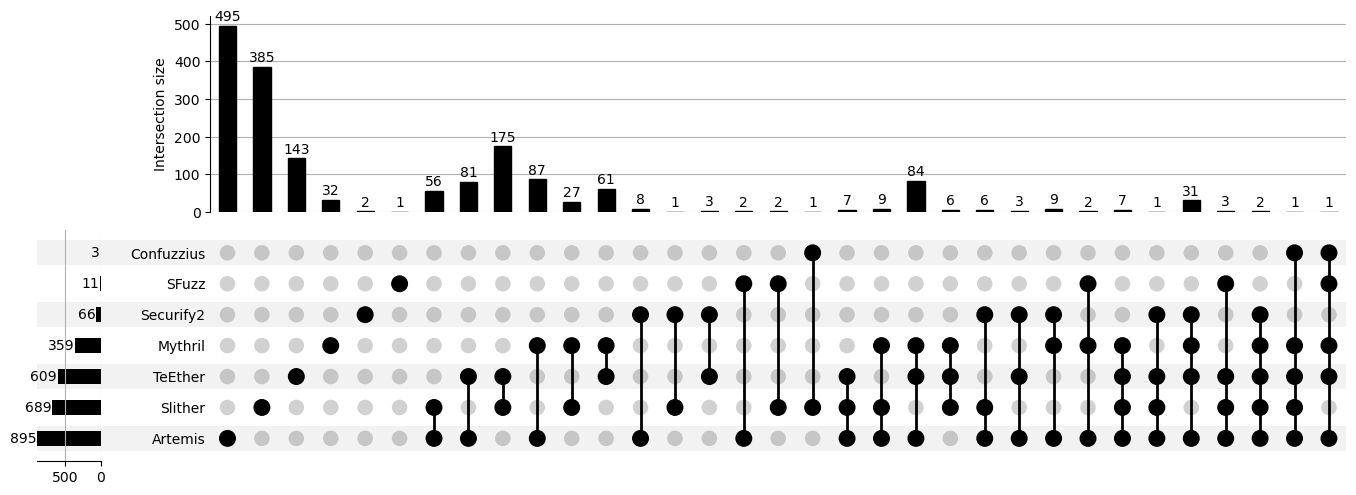}
\vspace{-1.4em}
\caption{Overlap of tools detecting the \textit{DelegateCall} vulnerability.
}\label{fig:source_dele}
\vspace{0.5em}
\end{figure}

\noindent\textbf{DelegateCall}
The \Cref{fig:source_dele} shows our evaluation results of six tools. As with the other vulnerability types, none of the tools agree on a single example. Further, there is a substantial overlap between Mythril and other tools -- specifically TeEther and Artemis.

\noindent\textbf{Summary} In summary, our quantitative analysis of all detectable vulnerabilities demonstrates a general lack of consensus among tools regarding the presence of any vulnerability. This indicates a significant gap between the reported detection capabilities and the actual data observed on-chain. The divergence in our detection results raises questions about the reliability of these tools in accurately identifying vulnerabilities in source code.

\subsection{Quantitative Analysis on \byteBig Dataset}
\label{sub:bytecode_eval}

\begin{table}[ht!]
	\centering
	\scalebox{1.0}{
\begin{tabular}{|c|c|c|c|c|c|c|c|c|c|c|c|c|c|c|c|c|c|}
\hline
\multirow{2}{*}{\thead{Metrics}} & \multicolumn{1}{l|}{ \thead{Static Analysis}} & \multicolumn{3}{l|}{\thead{Symbolic Execution}} \\ \cline{2-5} 
                   & Vandal     &  Maian   & Oyente   & Mythril  \\ \hline
Suicidal           & \CIRCLE    & \CIRCLE  & \Circle\ & \CIRCLE  \\ 
Reentrancy         & \CIRCLE    & \Circle\ & \CIRCLE  & \CIRCLE  \\ 
UncheckedCall      & \CIRCLE    & \Circle\ & \Circle\ & \CIRCLE  \\ 
txOrigin           & \CIRCLE    & \Circle\ & \Circle\ & \CIRCLE  \\ 
Timestamp          & \Circle\   & \Circle\ & \CIRCLE  & \CIRCLE  \\ \hline
\end{tabular}
}
   \caption{Overview of vulnerability scanners with detectable vulnerability types on bytecode.
   }   \label{tab:overview_tools_byte}
\end{table}

In this section, we evaluate four tools that can detect vulnerabilities in bytecode against our Bytecode dataset of~\numprint{4062844} unique bytecodes of \byteBig (cf.~\Cref{sub:bytecode_data}). 
We compare each tool's performance, detecting five different vulnerabilities: Reentrancy, Suicidal, Unchecked Call, Use of txOrigin, and Time Dependency.
We list an overview of the evaluated tools and their overlap for the detected vulnerability types in~\Cref{tab:overview_tools_byte}.

\begin{figure*}[htp]
\centering
\begin{minipage}{0.40\textwidth}
\includegraphics[width=\textwidth]{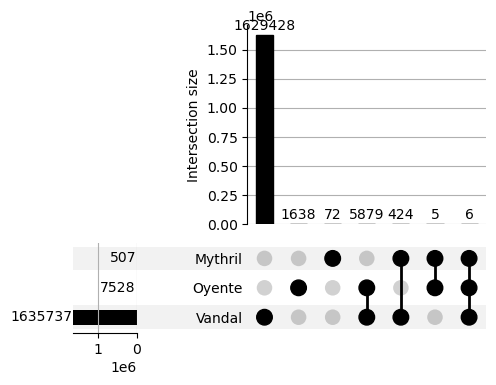}
\caption{Overlap of bytecode tools detecting the reentrancy vulnerability.
}\label{fig:byte_reentrancy}
\end{minipage}\hfill
\begin{minipage}{0.40\textwidth}
\includegraphics[width=\textwidth]{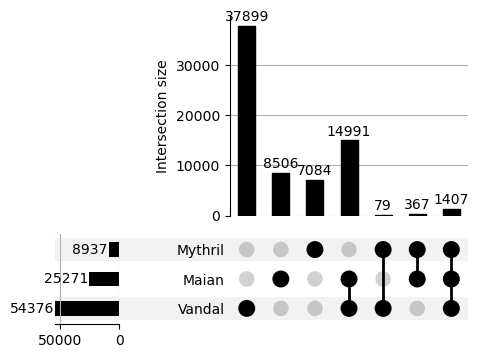}
\caption{Overlap of bytecode tools detecting the suicidal vulnerability.}\label{fig:byte_suicidal}
\end{minipage}
\end{figure*}

\noindent\textbf{Reentrancy}
As highlighted, a reentrancy vulnerability permits an attacker to deplete a smart contract's funds. 
\Cref{fig:byte_reentrancy} illustrates the intersections in detecting the Reentrancy bug among Mythril, Oyente, and Vandal. 
Vandal identifies the highest number of contracts vulnerable to reentrancy, leading to its greatest overlap with the other two tools. 
Conversely, Oyente and Mythril coincide in their identification on merely five contracts, while an overlap among all three tools occurs in just six distinct contracts.

\noindent\textbf{Suicidal} 
In \Cref{fig:byte_suicidal}, the overlap in identifying the Suicide vulnerability is depicted, a flaw that allows an attacker to terminate a smart contract. 
Mythril, Maian, and Vandal are capable of detecting this vulnerability, whereas Oyente is incapable of recognizing it in bytecode. 
Vandal, consistent with its performance in detecting other vulnerabilities, flags the most significant number of contracts as susceptible. 
Notably, the shared detection by all three tools is more substantial in this case compared to the Reentrancy vulnerability. 
Additionally, there is a significant overlap between Maian and Vandal, with both identifying the issue in approximately~\numprint{14991} contracts. 
Despite this, there remains a considerable disparity in the assessments across most contracts by the three analyzed tools.

\begin{figure*}[htp]
\centering
\begin{minipage}{0.30\textwidth}
\includegraphics[width=\textwidth]{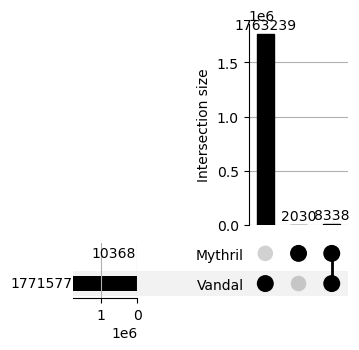}
\caption{Overlap of bytecode tools detecting the unchecked call vulnerability.
}\label{fig:byte_uncheck}
\end{minipage}\hfill
\begin{minipage}{0.30\textwidth}
\includegraphics[width=\textwidth]{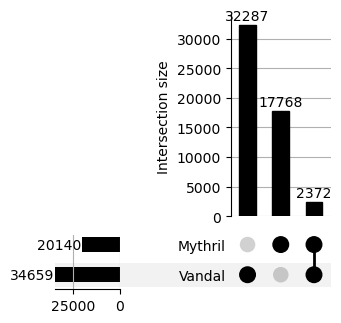}
\caption{Overlap of bytecode tools detecting the txOrigin vulnerability.}\label{fig:byte_origin}
\end{minipage}\hfill
\begin{minipage}{0.30\textwidth}
\includegraphics[width=\textwidth]{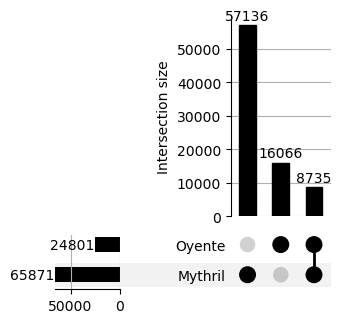}
\caption{Overlap of bytecode tools detecting the Time Dependency vulnerability.}\label{fig:byte_time}
\end{minipage}
\end{figure*}

\noindent\textbf{Unchecked Call}
Unchecked calls can be exploited by an attacker to induce undefined behavior in a smart contract.
This vulnerability is identifiable by Mythril and Vandal.
\Cref{fig:byte_uncheck} displays the intersection of these tools in detecting the issue within bytecode.
We see Vandal's tendency to report excessively, in this case identifying \numprint{1771577} instances as vulnerable, which is over one-fourth of all contracts. 
Further, there is a significant degree of overlap with Mythril's findings.
Nevertheless, this overlap constitutes less than~\SI{0.5}{\percent} of the total detections of Vandal.

\noindent\textbf{TxOrigin}
The use of txOrigin gives an attacker the potential to bypass authorization checks in a smart contract. 
The overlap between Mythril and Vandal in detecting this issue is illustrated in~\Cref{fig:byte_origin}. 
The two tools agree on the vulnerability in only~\numprint{2372} contracts, representing less than~\SI{5}{\percent} of the total contracts in our dataset. 
Collectively, these tools have flagged~\SI{1.3}{\percent} of contracts, indicating a possible overestimation of this vulnerability when all tools are utilized together.

\noindent\textbf{Time Dependency}
The intersection of Oyente and Mythril in detecting the Time Dependency vulnerability is depicted in \Cref{fig:byte_time}. 
These tools agree on the vulnerability in~\numprint{8735} contracts. However, they diverge in their assessments for over~\SI{90}{\percent} of the contracts they flagged, indicating a significant discrepancy in their detection capabilities for this particular vulnerability.

\noindent\textbf{Summary} Overall, there is a parallel pattern observed with bytecode analysis tools compared to those analyzing source code. 
Consistently, there's a notable disagreement among the tools regarding the majority of contracts and types of vulnerabilities. 
This pattern suggests that these tools may not be highly effective on a larger scale. 
Although this quantitative analysis doesn't fully capture the exact performance of the bytecode tools, the evident discrepancies among various tools are clear.

\subsection{Source Code vs. Bytecode}
\label{sub:source_byte_eval}
We consider two vulnerability scanners, Mythril and Maian, that allow the input of source code and bytecode.
In this section, we evaluate the performance difference between these scanners and see whether they have a performance difference being exposed to source code or bytecode.
This analysis is based on our quantitative datasets.

\noindent\textbf{Maian} flags four smart contracts as Suicidal in source code and \numprint{93} in bytecode, overlapping in three cases. It also finds \numprint{80} contracts with locked ether in source code and \numprint{3223} in bytecode, with \numprint{66} common detections. Generally, Maian shows a higher positive rate in bytecode analysis.

\noindent\textbf{Mythril} shows no common detections between source code and bytecode for the DelegateCall and Reentrancy vulnerabilities and only one shared detection for the Suicide vulnerability. It finds an overlap of~\numprint{35} in Arithmetic Bugs, \numprint{76} in txOrigin, and a notable \numprint{1242} in Time Dependency.
Contrary to Maian, Mythril generally presents a higher positive detection rate in source codes compared to bytecodes.

\subsection{Qualitative Analysis based on \GTReentrancyNoSpace Dataset}
\label{sub:reent_eval}
In this section, we evaluate both source code and bytecode analysis tools using our meticulously curated reentrancy dataset \GTReentrancyNoSpace, as detailed in \Cref{sub:reentrancy_data}. 
As mentioned, this dataset is exclusively annotated for the reentrancy vulnerability, which is categorized into three subtypes: call, send, and transfer. 

Our analysis covers two distinct approaches: the first includes only the 'call' subtype under the umbrella of reentrancy, while the second encompasses all three subtypes. 
This dual approach aids in determining the specific type of reentrancy each tool is geared to identify, especially since the tools do not explicitly state their focus on certain subtypes.

\noindent\textbf{Analysis of Source Code-based Tools} We present our findings for the call subtype in~\Cref{tab:source_gt_call_reent}, which displays evaluation results for nine vulnerability scanners. 
In contrast to the quantitative analysis, we do not consider samples if the applied vulnerability scanner does not finish the evaluation.

Securify2 emerged as the most effective tool with an F1-Score of 38\%, but it was only able to test \numprint{662} instances. The key observation here is the generally poor detection rate for the call subtype across all tools.

\begin{table}[ht!]
	\centering
	\scalebox{1.0}{
\begin{tabular}{|l|c|c|c|c|c|c|}
\hline
Tools            & TN & FP & FN & TP & F1 & Acc \\ \hline
Artemis          & \numprint{5174}       & \numprint{95}        & \numprint{358}      & \numprint{71}        & \numprint{0.24}        & \numprint{0.92} \\ 
Osiris           & \numprint{1727}       & \numprint{81}        & \numprint{112}      & \numprint{49}        & \numprint{0.34}        & \numprint{0.90} \\ 
Confuzzius       & \numprint{2956}       & \numprint{21}        & \numprint{159}      & \numprint{30}        & \numprint{0.25}        & \numprint{0.94} \\ 
Smartian         & \numprint{2279}       & \numprint{3}         & \numprint{168}      & \numprint{10}        & \numprint{0.10}        & \numprint{0.93} \\ 
Mythril          & \numprint{4069}       & \numprint{398}       & \numprint{298}      & \numprint{83}        & \numprint{0.19}        & \numprint{0.86} \\ 
sFuzz            & \numprint{617}        & \numprint{24}        & \numprint{34}       & \numprint{7}         & \numprint{0.19}        & \numprint{0.91} \\ 
Oyente           & \numprint{2854}       & \numprint{0}         & \numprint{169}      & \numprint{1}         & \numprint{0.01}        & \numprint{0.94} \\ 
Slither          & \numprint{4469}       & \numprint{1483}      & \numprint{41}       & \numprint{411}       & \numprint{0.35}        & \numprint{0.76} \\ 
Securify2        & \numprint{603}        & \numprint{41}        & \numprint{4}        & \numprint{14}        & \numprint{0.38}        & \numprint{0.93} \\ \hline
\end{tabular}
}
   \caption{Results of Ground Truth call-subtype Reentrancy dataset with source code tools.}   \label{tab:source_gt_call_reent}
\end{table}

The results for all reentrancy subtypes are presented in~\Cref{tab:source_gt_all_reent}. Here, we observe a general decline in detection effectiveness among most tools. However, three tools -- Slither, Securify2, and Mythril -- show improved performance, suggesting their emphasis on all three reentrancy bug subtypes.

\begin{table}[ht!]
	\centering
	\scalebox{1.0}{
\begin{tabular}{|l|c|c|c|c|c|c|}
\hline
Tools            & TN & FP & FN & TP & F1 & Acc \\ \hline
Artemis          & \numprint{4378}       & \numprint{53}        & \numprint{1154}      & \numprint{113}        & \numprint{0.16}        & \numprint{0.79} \\ 
Osiris           & \numprint{1257}       & \numprint{46}        & \numprint{582}       & \numprint{84}         & \numprint{0.21}        & \numprint{0.68} \\ 
Confuzzius       & \numprint{2484}       & \numprint{14}        & \numprint{631}       & \numprint{37}         & \numprint{0.10}        & \numprint{0.80} \\ 
Smartian         & \numprint{1931}       & \numprint{2}         & \numprint{516}       & \numprint{11}         & \numprint{0.04}        & \numprint{0.79} \\ 
Mythril          & \numprint{3517}       & \numprint{298}       & \numprint{850}       & \numprint{183}        & \numprint{0.24}        & \numprint{0.76} \\ 
sFuzz            & \numprint{411}        & \numprint{17}        & \numprint{240}       & \numprint{14}         & \numprint{0.10}        & \numprint{0.62} \\ 
Oyente           & \numprint{2279}       & \numprint{0}         & \numprint{744}       & \numprint{1}          & \numprint{0.00}        & \numprint{0.75} \\ 
Slither          & \numprint{4324}       & \numprint{704}       & \numprint{186}       & \numprint{1190}       & \numprint{0.73}        & \numprint{0.86} \\ 
Securify2        & \numprint{592}        & \numprint{24}        & \numprint{15}        & \numprint{31}         & \numprint{0.61}        & \numprint{0.94} \\ \hline
\end{tabular}
}
   \caption{Results of Ground Truth all subtypes Reentrancy dataset with source code tools.}   \label{tab:source_gt_all_reent}
\end{table}

This analysis reveals varied interpretations of 'reentrancy' across tools. Most prioritize the call subtype, but some also regard send and transfer subtypes as important. This variance in definitions hinders direct performance comparisons between tools.

\noindent\textbf{Analysis of Bytecode-based Tools}
As we already explained in~\Cref{sub:reentrancy_data}, 
we do not compile smart contracts of our dataset.
Instead, we use their addresses to download the bytecode from the blockchain.
This allows us to analyze the deployed smart contract version without introducing discrepancies due to compiler versions and optimizations.

The outcomes of bytecode tools for the call-subtype reentrancy are presented in \Cref{tab:byte_gt_call_reent}. 
It's observed once more that bytecode tools are capable of analyzing a greater number of contracts compared to their source code counterparts. 
Additionally, it's notable that Mythril's performance is less effective on bytecode than on source code, which is understandable considering the richer information available in the source code. 
Vandal, despite achieving the highest F1-score among the tools, falls behind in terms of accuracy compared to the others. 
Interestingly, Oyente demonstrates improved performance on bytecode over its analysis of the same data in source code. 
Overall, the effectiveness of these tools in analyzing bytecode for call-subtype reentrancy is not particularly impressive.

\begin{table}[ht!]
	\centering
	\scalebox{1.0}{
\begin{tabular}{|l|c|c|c|c|c|c|}
\hline
Tools            & TN & FP & FN & TP & F1 & Acc \\ \hline
Vandal          & \numprint{3662}       & \numprint{2292}     & \numprint{107}      & \numprint{345}      & \numprint{0.22}       & \numprint{0.63}\\ 
Oyente          & \numprint{5900}       & \numprint{54}       & \numprint{412}      & \numprint{40}       & \numprint{0.15}       & \numprint{0.93}\\ 
Mythril         & \numprint{5953}       & \numprint{1}        & \numprint{452}      & \numprint{0}        & \numprint{0.00}       & \numprint{0.93}\\ \hline
\end{tabular}
}
   \caption{Results of Ground Truth call-subtype Reentrancy dataset with bytecode tools.}   \label{tab:byte_gt_call_reent}
\end{table}

The findings for all reentrancy subtypes are presented in \Cref{tab:byte_gt_all_reent}. 
In this comparison, both Oyente and Mythril exhibit diminished performance. 
On the other hand, Vandal improves its F1-score significantly, jumping from~\SI{22}{\percent} to~\SI{40}{\percent}. 
However, it's important to note that Vandal's accuracy decreases by one percent.

\begin{table}[ht!]
	\centering
	\scalebox{1.0}{
\begin{tabular}{|l|c|c|c|c|c|c|}
\hline
Tools            & TN & FP & FN & TP & F1 & Acc \\ \hline
Vandal          & \numprint{3194}       & \numprint{1836}     & \numprint{575}      & \numprint{801}      & \numprint{0.40}       & \numprint{0.62}\\ 
Oyente          & \numprint{4995}       & \numprint{35}       & \numprint{1317}      & \numprint{59}       & \numprint{0.08}       & \numprint{0.79}\\ 
Mythril         & \numprint{5029}       & \numprint{1}        & \numprint{1376}      & \numprint{0}        & \numprint{0.00}       & \numprint{0.79}\\ \hline
\end{tabular}
}
   \caption{Results of Ground Truth all subtypes Reentrancy dataset with bytecode tools.}   \label{tab:byte_gt_all_reent}
\end{table}

\noindent\textbf{Summary.} Ultimately, Slither had the best F1-Score overall on our handcrafted reentrancy dataset with~\SI{73}{\percent} for the source code-based analysis.
Although bytecode analysis tools demonstrated a higher completion rate, their detection efficacy was notably lower compared to when the vulnerability scanners were applied directly to the source code.
A key insight from this study is the significant influence that vulnerability scanners' definitions of a vulnerability have on the performance of various tools on the same dataset. 

Our investigation focused on the nuances of the reentrancy bug, a well-recognized type of vulnerability. 
However, it's plausible that similar discrepancies in the definition of vulnerabilities in different studies could also affect the detection of other types of vulnerabilities.

\subsection{Qualitative Analysis based on \GTAudits Dataset}
\label{sub:audit_eval}
In our Audit dataset, we ensure that all smart contracts are compiled using the required compiler version, facilitating the analysis of their bytecode. 
On the other hand, when it comes to source code-based analysis, we encounter limitations.  
Due to the small number of vulnerable contracts and the inability of tools to analyze them without modifications to either the project's source code or the vulnerability scanners, we couldn't conduct source code-based analysis effectively. 
Consequently, our focus in this section is primarily on the analysis of bytecode.

\noindent\textbf{Reentrancy} We show the results of the reentrancy bug detection in~\Cref{tab:byte_audit_reent}.
Like in the case of our self-labeled dataset \GTReentrancy (cf.~\Cref{sub:reentrancy_data}), all three analyzed tools show low detection performance.

\begin{table}[ht!]
	\centering
	\scalebox{1.0}{
\begin{tabular}{|l|c|c|c|c|c|c|}
\hline
Tools            & TN & FP & FN & TP & F1 & Acc \\ \hline
Vandal          & \numprint{87}        & \numprint{31}       & \numprint{31}      & \numprint{36}       & \numprint{0.40}       & \numprint{0.53}\\ 
Oyente          & \numprint{165}       & \numprint{0}        & \numprint{67}      & \numprint{0}        & \numprint{0.00}       & \numprint{0.71}\\ 
Mythril         & \numprint{163}       & \numprint{2}        & \numprint{67}      & \numprint{0}        & \numprint{0.00}       & \numprint{0.70}\\ \hline
\end{tabular}
}
   \caption{Results of Audit dataset for Reentrancy with bytecode tools.}   \label{tab:byte_audit_reent}
\end{table}

\noindent\textbf{Other Vulnerabilities.} Our additional findings are detailed in~\Cref{tab:byte_audit_unchecked}. 
In this table, we present the tools used and the corresponding SWC identifiers for each vulnerability, as referenced in \Cref{subsec:vulns}.
These results further highlight the limited effectiveness of all involved tools in detecting various vulnerabilities.

\begin{table}[ht!]
	\centering
	\scalebox{0.87}{
\begin{tabular}{|l|c|c|c|c|c|c|c|}
\hline
Tools            & TN & FP & FN & TP & F1 & Acc & Vulnerability (SWC)\\ \hline
Mythril       & \numprint{187}       & \numprint{18}       & \numprint{25}       & \numprint{2}       & \numprint{0.09}       & \numprint{0.81} & Int Over/Underflow (101)\\ 
Vandal          & \numprint{52}       & \numprint{28}       & \numprint{90}       & \numprint{62}       & \numprint{0.51}       & \numprint{0.49}& Unchecked Call (104)\\ 
Oyente         & \numprint{216}       & \numprint{0}       & \numprint{16}       & \numprint{0}       & \numprint{0.00}       & \numprint{0.93}& ToD (114)\\ 
Mythril        & \numprint{202}       & \numprint{18}       & \numprint{9}       & \numprint{3}       & \numprint{0.18}       & \numprint{0.88}& Time Dependency (116)\\ 
Oyente         & \numprint{220}       & \numprint{0}       & \numprint{12}       & \numprint{0}       & \numprint{0.00}       & \numprint{0.95}& Time Dependency (116)\\ \hline
\end{tabular}
}
   \caption{Results of Audit dataset for different vulnerabilities with bytecode tools.}   \label{tab:byte_audit_unchecked}
\end{table}

\subsection{Scanning Robustness}
\label{sub:robustness_eval}

In this section, we assess the robustness of the vulnerability scanners.
Given that some scanners are unable to complete the analysis of certain smart contracts, our focus is on the frequency of successful analyses.
This is in the interest of smart contract developers since an unfinished scan leaves them with some degree of uncertainty.

\begin{figure}[ht!]
\centering
\includegraphics[width=0.8\columnwidth]{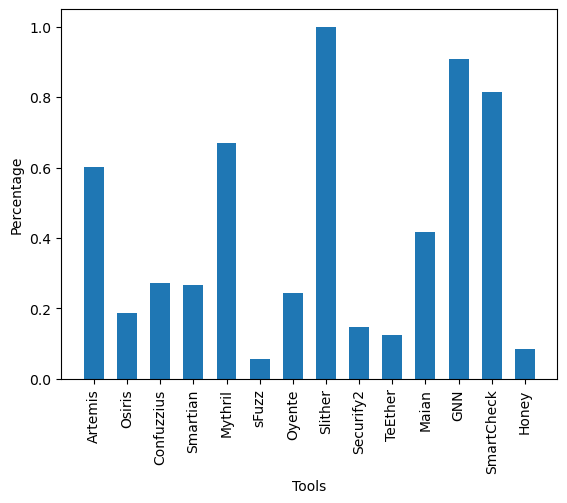}
\vspace{-1.0em}
\caption{Robustness of Tool Detection
}\label{fig:robust_source}
\end{figure}

\Cref{fig:robust_source} shows the percentage of how many contracts were successfully scanned by the different source code-based tools.
We can see that only five tools (Artemis, Mythril, Slither, GNN, and Smartcheck) were able to at least check half of the smart contracts in our dataset.
Otherwise, the tools quite often were not able to complete their analysis.

We attribute these disheartening results to several factors.
First, most tools base their analysis on specific compiler versions of the Solidity compiler.
Since the area of smart contracts is a rather fast-growing field, compiler versions are regularly updated and at some points introduce breaking changes.
A further concern involves reliance on third-party tools, like constraint-solving algorithms. These tools frequently receive updates, which might lead to compatibility issues with existing versions of the tools or Solidity.
Further, some smart contracts are written in another language. However, most tools are only able to analyze Solidity code.
Generally, these effects can be attributed to the effect of software aging. Most pressing is the issue of dependency on specific compiler versions, which makes most of the tooling dependent on a fast-changing piece of code.

\section{Discussion} 
\label{sec:discussion}

In this section, we engage in a focused discussion about our study's outcomes. 
We first analyze why the tools used show discrepancies in their results, considering their different methodologies and biases. 
Next, we debate if using more tools leads to improved vulnerability detection, balancing the pros and cons of diverse methodologies against potential complexities. 
We then evaluate the ease of setup and use for these tools, considering their user-friendliness and technical requirements. 
Lastly, we assess the commonness of vulnerabilities in smart contracts, understanding their prevalence and impact on the security and efficacy of these contracts. 
This discussion aims to provide a comprehensive understanding of our findings and their implications in the practical use of smart contracts vulnerability scanners.

\subsection{Disagreement of Tools}

The disagreement among tools is largely attributed to the varying times at which they were developed, leading to three key issues which we discuss below. 

\noindent\textbf{Compiler Version Changes} Tools, particularly those analyzing source code, are often designed for specific versions of the Solidity compiler. As the compiler evolves, these tools may not be updated promptly to accommodate new versions or resolve potential breaking changes.

\noindent\textbf{Evolution of Smart Contract Programming} The programming practices in smart contracts can shift over time. For instance, the use of `send`/`transfer` calls was initially common in smart contracts. However, changes in Ethereum, like the gas limit alterations for these calls, have led to programmers being advised against their use.

\noindent\textbf{New Developments and Attack Variations} The emergence of new developments and variations in attack strategies can alter the landscape of vulnerabilities. An example is the introduction of built-in defenses against integer overflows and underflows in the compiler.

Additionally, vulnerability definition can differ across tools, influenced by the authors' perspectives and interpretations of what constitutes a vulnerability. In some cases, vulnerabilities might be part of a multi-step exploit process.

Overall, the significant disagreement among tools is concerning. It suggests that their detection capabilities are not yet robust enough to eliminate bugs effectively, a conclusion further reinforced by the ongoing exploits occurring on the blockchain.

\subsection{Usage of Multiple Tools}

Using a combination of tools to identify vulnerabilities in a smart contract might seem like a promising strategy. The differing results from various tools suggest that their combined use could potentially offer a more thorough analysis. However, this method has its drawbacks, primarily the increase in positive detections, which includes false positives. This means that more extensive portions of the smart contract would require examination.

Moreover, the challenge lies in selecting an appropriate mix of tools. There's no definitive or universally effective combination that guarantees the successful detection of all vulnerabilities. Each tool has its own strengths, weaknesses, and focus areas, and their efficiency can vary depending on the contract's specific features and vulnerabilities.

In summary, there is no foolproof or 'silver bullet' method for detecting all vulnerabilities in smart contracts. While using multiple tools might enhance the breadth of analysis, it does not necessarily lead to flawless detection and can complicate the assessment process.

\subsection{Difficulty of Usage}

The implementation ease and use of smart contract vulnerability detection tools vary widely. Some tools are easily operated with a Docker command, while others require complex dependency resolutions, especially Java-based tools using Maven. Running these tools often necessitates specific configurations and parameters, adding complexity and reducing practicality for developers.

Further, we discuss a range of smart contract analysis tools, which were not included in our analysis due to their source code being inaccessible. These tools encompass EthPloit~\cite{zhang2020ethploit}, Harvey~\cite{wustholz2020harvey}, ReGuard~\cite{liu2018reguard}, Solar~\cite{li2018detecting}, SmartScopy~\cite{feng2019precise}, SESCon~\cite{ali2021sescon} and sCompile~\cite{chang2019scompile}, Ether* \ S-gram~\cite{liu2018s}, Sereum~\cite{rodler2018sereum}, Easyflow~\cite{easyflow_paper}, Zeus~\cite{zeustool_paper}, Ethainter~\cite{ethainter_paper} and the framework described in~\cite{reentrancytool_paper}.  
Similar considerations apply to SACS~\cite{zhou2018security} and Sailfish~\cite{bose2022sailfish}. Despite assurances of making their code open source, these tools remain unavailable as of the time this paper was written. 

On a different note, the tools from which ESCORT~\cite{escort} and other ML approaches derive their learning are available and are consequently incorporated into our analysis.

In our analysis, we intended to include a wider range of tools but faced challenges due to issues like complex setups and outdated functionalities. Tools such as ContractFuzzer~\cite{jiang2018contractfuzzer}, EtherRacer~\cite{kolluri2019exploiting}, SODA~\cite{chen2020soda}, and SecurifyV1~\cite{securify_paper} were non-functional because of outdated or unmaintained repositories, though we could include the updated SecurifyV2.

NeuCheck~\cite{lu2021neucheck} lacked clear setup instructions, making it unsuitable for our study. Tools like eThor~\cite{ethor_paper} and EthBMC~\cite{ethbmc_paper}, requiring intensive constraint solving, had prohibitive execution times for large-scale analysis.

Lastly, we excluded tools such as Manticore~\cite{mossberg2019manticore} and MadMax~\cite{madmax_paper}, which focus on metrics like code coverage and gas issues, respectively, not aligning with our vulnerability detection objective.
\section{Related Work} 
\label{sec:related_work}

The related work concerning analysis of smart contract vulnerability detection tools can be categorized into two categories: Theoretical and practical analysis.

\vspace{0.2cm}\noindent\textbf{Theoretical Analysis}
Several studies including~\cite{tang2021vulnerabilities,rameder2022review,zhou2022state,praitheeshan2019security,sayeed2020smart,kushwaha2022systematic} have conducted theoretical analyses of smart contract detection tools and vulnerabilities in Ethereum smart contracts. These studies primarily focus on categorizing vulnerabilities, listing available tools, and discussing their properties without performing actual evaluations or comparisons.
In summary, these studies provide theoretical insights into smart contract vulnerabilities and detection tools, but they do not offer practical assessments of how well these tools perform in real-world scenarios.

\vspace{0.2cm}\noindent\textbf{Practical Analysis}
The SmartBugs framework~\cite{ferreira2020smartbugs} compares ten smart contract analysis tools using a dataset of 143 annotated contracts. 

The Solidify framework~\cite{ghaleb2020effective} assesses six static analysis tools on 50 contracts injected with 9,369 bugs, with each vulnerability randomly represented in the code.

Ren et al.~\cite{ren2021empirical} analyzed nine tools out of the three categories: static analysis, symbolic execution, and dynamic fuzzing. Their study utilized a dataset that encompassed real-world contracts, manually injected bugs, and verified vulnerable contracts, culminating in a total of 46,186 unique contracts, of which 214 were confirmed as vulnerable. In contrast to our research, their investigation was solely concentrated on reentrancy vulnerabilities in source codes.

Kushwaha et al.~\cite{kushwaha2022ethereum} performed a theoretical comparison of 86 analysis tools, as documented in 145 research papers. From this survey, they selected 16 tools for analysis, primarily focusing on categories such as symbolic execution and constraint solving. Their analysis was based on a relatively limited dataset comprising only 30 contracts, tagged with five specific vulnerabilities.

Dika and Nowostawski~\cite{dika2018security} provided insights into four tools using a dataset of 45 contracts, split between 21 clean and 24 vulnerable. 

He et al.~\cite{he2020smart} focused on random number vulnerabilities in Fomo3d-like games and discussed three auditing tools for smart contract security.

Peng et al.~\cite{qian2022smart} presented an overview of 29 smart contract analysis tools, assessing their language support, analysis methods, and detectable vulnerabilities. They compared five tools using a dataset of 300 randomly collected smart contracts from Etherscan.

While previous studies, provide practical evaluations, they are constrained in scope, primarily focusing on a narrow selection of tools or utilizing relatively small datasets for comparison. Notably, none of these surveys address the intersection of tools, a significant oversight in the related literature. This gap is crucial as it reveals considerable discrepancies in results across almost every tool, highlighting the importance of comprehensive and varied datasets for robust tool evaluation.
\section{Conclusion} 
\label{sec:conclusion}

To conclude, our extensive analysis, encompassing millions of smart contracts with both source codes and bytecodes, including those that are manually labeled, highlights a clear finding: there is substantial scope for enhancement in the realm of smart contract security. 
This study underscores the ongoing and complex nature of the challenge of detecting vulnerabilities effectively.

\bibliographystyle{IEEEtran}
\bibliography{refs}
\appendix
\section{Appendix}
\label{sec:appendix}

\subsection{Additional Figures}
\begin{sidewaysfigure}

\centering
\includegraphics[width=1.0\textwidth]{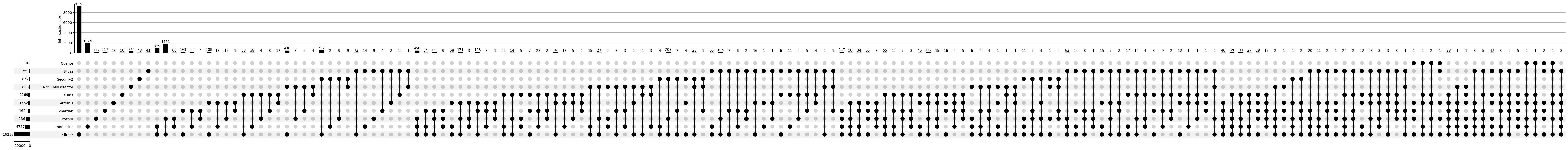}
\caption{Overlap of tools detecting the \textit{Time Dependency} vulnerability in source code.
}\label{fig:source_time}
\vspace{0.5em}
\end{sidewaysfigure}

\Cref{fig:source_time} shows our analysis of the Time Dependency vulnerability on our source code dataset~\sourceBigNoSpace.

\end{document}